# Color of Copper/Copper oxide

Su Jae Kim[1,†], Seonghoon Kim[2,†], Jegon Lee[3,†], Youngjae Jo[4], Yu-Seong Seo[3], Myounghoon Lee[3], Yousil Lee[1], Chae Ryong Cho[5], Jong-pil Kim[6], Miyeon Cheon[1], Jungseek Hwang[3], Yong In Kim[7,8], Young-Hoon Kim[7,8], Young-Min Kim[7,8], Aloysius Soon[9], Myunghwan Choi[10], Woo Seok Choi[3*], Se-Young Jeong[7,11*], & Young Hee Lee[3,7,8*]

[1]Crystal Bank Research Institute, Pusan National University, Busan 46241, Republic of Korea.
[2]Research institute of basic science, Seoul National University, Seoul 08826, Republic of Korea.
[3]Department of Physics, Sungkyunkwan University, Suwon 16419, Republic of Korea.
[4]Center for Neurscience Imaging Research, Sungkyunkwan University, Suwon 16419, Republic of Korea.
[5]Department of Nanoenergy Engineering, Pusan National University, Busan 46241, Republic of Korea.
[6] Division of High-Tech Materials Research, Busan Center, Korea Basic Science Institute, Busan 46742, Republic of Korea.
[7]Center for Integrated Nanostructure Physics, Institute for Basic Science, Sungkyunkwan University Suwon 16419, Republic of Korea.
[8]Department of Energy Science, Sungkyunkwan University, Suwon 16419, Republic of Korea.
[9]Center for Artificial Synesthesia Materials Discovery and Department of Materials Science and Engineering, Yonsei University, Seoul 03722, Republic of Korea
[10]School of Biological Sciences, Seoul National University, Seoul 08826, Republic of Korea.
[11]Department of Optics and Mechatronics Engineering, Pusan National University, Busan 46241, Republic of Korea.

* e-mail: choiws@skku.edu, syjeong@pusan.ac.kr, leeyoung@skku.edu

# Abstract

Stochastic inhomogeneous oxidation is an inherent characteristic of copper (Cu), often hindering color tuning and bandgap engineering of oxides. Coherent control of the interface between metal and metal oxide remains unresolved. We demonstrate coherent propagation of an oxidation front in single-crystal Cu thin film to achieve a full-color spectrum for Cu by precisely controlling its oxide-layer thickness. Grain boundary-free and atomically flat films prepared by atomic-sputtering epitaxy allow tailoring of the oxide layer with an abrupt interface via heat treatment with a suppressed temperature gradient. Color tuning of nearly full-color RGB indices is realized by precise control of oxide-layer thickness; our samples covered ~50.4% of the sRGB color space. The color of copper/copper oxide is realized by the reconstruction of the quantitative yield color from oxide "pigment" (complex dielectric functions of $Cu_2O$) and light-layer interference (reflectance spectra obtained from the Fresnel equations) to produce structural color. We further demonstrate laser-oxide lithography with micron-scale linewidth and depth through local phase transformation to oxides embedded in the metal, providing spacing necessary for semiconducting transport and optoelectronics functionality.

# Introduction

Surface oxidation of copper (Cu), one of the oldest problems in metallurgy, occurs naturally when Cu is exposed to air. The oxidation depends on the imposed environmental conditions.[1-3] The tensor relationship between control parameters and oxidation has not been addressed thus far, because the propagation of oxidation occurs randomly at the surface with a high density of low coordinated surface sites (preferentially along grain boundaries). Oxidation is neither prevented nor systematically controlled along a uniaxial direction. Thus, systematic control over surface oxidation is necessary to take full advantage of the properties of metals.

Color modulation of metals has been attempted by exploitation of electrochromism, laser coloration using marking, piezochromism, and plasmonic effects.[4-7] Highly porous thin films on metal substrates with ultra-thin, finely tuned optical coatings offer color purity enhancement.[8-10] However, Cu and its alloys become tarnished and corrode under the ambient conditions often involved in antimicrobial applications on various touch surfaces in healthcare facilities.[11,12] Another strategy to obtain wide color selectivity in a metal film is the construction of sophisticated nanostructures to realize various colors by means of polarization conversion.[13] Despite numerous attempts to modulate color by oxidation and nanostructuring efforts,[14-16] the complexity associated with conversion of the Cu lattice into an oxide remains an obstacle for coherent control of the interface between metal and metal oxide, which is necessary to obtain a full, well-defined color spectrum.

In this report, we present a breakthrough in the surface oxidation of Cu, using a grain boundary-free, ultra-flat single-crystal Cu thin film (SCCF) prepared by atomic-sputtering epitaxy (ASE). Inhomogeneous oxidation in the SCCF was highly suppressed by introduction of a treatment to minimize the temperature gradient in the film, resulting in the production of a full-color spectrum by precise control of the oxide-layer thickness. This approach is further extended to localized oxidation by laser-oxide lithography for photonic-electronic applications.

**Results & discussion**

Figure 1a shows a schematic diagram of ASE, in which all internal electrical circuits are replaced by single-crystal Cu wires instead of commercial Cu wires; the vibration due to ambient noise is highly suppressed by an anti-vibration system (Experimental Section/Methods). However, atomic sputter epitaxy (ASE) aims to realize atomically flat surfaces by stacking atom by atom. Hence, even minute vibration could significantly disturb initial nucleation and lateral growth, especially the coherent coplanar merging of the nuclei. We verified that precise signal transduction and cancellation of electrical interference through the use of grain boundary-free wires were essential for acquisition of high-quality Cu thin films. Indeed, ASE resulted in grain boundary-free, single-crystal Cu thin films (Figure 1b top panel, SCCF) with a root mean square (RMS) roughness of 0.25 nm, comparable to the thickness of a single atomic layer along the (111) crystallographic direction of Cu (Figure S1, Supporting Information). To highlight the single-crystal nature of our SCCF, we show the surface quality of a partially improved Cu thin film grown using single-crystal target-assisted sputtering (Figure 1b middle panel, intermediate),[17] as well as a general polycrystalline Cu thin film (PCCF) grown using a conventional sputtering system [Figure 1b (bottom panel) and Figure S2, Supporting Information]. Numerous grains and grain boundaries were evident in scanning electron microscopy (SEM) images (left), with large RMS roughness values of 3.98 and 11.29 nm from atomic force microscopy (AFM) images (right); in contrast, the SCCF had an RMS roughness of 0.25 nm.

A uniform, controllable oxide layer was obtained from the SCCF in a separate heating furnace. We annealed both SCCF and PCCF films at 330 °C for 1 min under a mixed-gas atmosphere of Ar (83%) and $O_2$ (17%). The SCCF exhibited an oxide layer with a highly crystalline $Cu_2O$ layer and nanometer-scale abrupt interfaces. Figure 1 shows cross-sectional high-resolution (scanning) transmission electron microscopy ((S)TEM) images of the oxide layers of the SCCF (Figures 1c and 1d) and PCCF (Figure 1e); the interface between the crystalline $Cu_2O$ and SCCF layers was found to be atomically sharp (abrupt) showing the single-crystallographic orientation by the fast Fourier transform (FFT) patterns (inset). Comparison of interplanar spacing profiles obtained along the out-of-plane direction of HAADF-STEM images between $Cu_2O$ and the Cu film revealed that the Cu film possessed the layer spacing of the (111) stacking plane (dCu(111) = 0.21 nm); the oxidized area exhibited the larger lattice spacing of $Cu_2O$ (d$Cu_2$O(111) = 0.25 nm). In contrast, the conventional PCCF exhibited a rough surface and interface with blurry FFT patterns (Figure 1e). The interface misfit strain is steeply relaxed over about 2 nm or less by the presence of geometrical misfit dislocations and the residual strain is gradually relieved over about 4 nm into the $Cu_2O$ layer. (Figures. S3 and S4, Supporting information). We emphasize that coherent oxidation, with a highly crystalline $Cu_2O$ layer and nanometer-scale abrupt interfaces, is critical for realization of homogeneous, genuine color in Cu films. This is achieved by minimization of the temperature gradient in the Cu film during heat treatment. For this purpose, we designed a double heating system equipped with an interior preheating furnace, in which the temperature was controlled to within ± 0.1 °C (Experimental Section/Methods).

By dramatically improving the interface quality with controllable oxidation in the SCCF, we realized a wide-color spectrum through exclusive use of simple heat treatment via temperature-gradient minimization. Figure 2a shows a photograph of representative SCCF samples with systematically controlled oxide-layer thicknesses. This vivid representation of color, which was not previously achieved, emphasizes "color" as an indicator of the surface/interface quality of the $Cu_2O$/Cu heterostructure; thus, it is an indicator of controlled oxidation. The color wheel (Figure 2b) shows representative colors of Cu films constructed from photographic images of actual samples (Figure S5, Supporting Information).[18,19] The range of colors achieved by our samples (>300) is mapped as a Commission Internationale de L'Eclairage (CIE) xy chromaticity diagram (Figure 2c).[20] Notably, colors from the SCCF cover ~50.4% of the area of the sRGB color space of digitally available colors represented by the grey triangle, which represents enhancement of >250% in color coverage relative to a recent study regarding plasmonic color generation (18.4%).[20] The photographic images were also obtained using near-normal geometry. The spectra of incident lights used for the images correspond to AM1.5. The sample colors change slightly according to the different angles of view (Figure S6, Supporting Information).

The underlying mechanism for the emergence of various colors can be explained in terms of the multiple reflections that occur at the oxidized film surface and interface between $Cu_2O$ and Cu,[21,22] as shown schematically in Figure 3a. The reflected light depends strongly on the $Cu_2O$ thickness. The change in color of incident white light that occurs upon reflection, is determined by the dielectric functions of the material. First, the dielectric functions of our $Cu_2O$ layer are consistently obtained by spectroscopic ellipsometry measurements (Figure S7 and S8, Supporting Information).[21-25] Based on the obtained dielectric functions of each layer, distinct reflectance spectra in the near-normal incident geometry are simulated for specific CuO/ $Cu_2O$ thicknesses, using the Fresnel equations. Figure 3b shows the simulated reflectance spectra (dotted lines. The simulated spectra are remarkably similar to experimental reflectance spectra (solid lines) of CuO/$Cu_2O$/SCCF thin films annealed at 330 °C for various durations with $Cu_2O$ layer thicknesses of 0, 28, 35, and 60 nm, thus validating the thickness modulation of multiple reflections and realization of the color variation (Figure S9, Supporting Information).

We investigated the critical variables for homogeneous color with a given initial Cu thickness (d) by controlling the annealing temperature (T) and annealing time (t). The systematic color variation from brown to purple was obtained with the corresponding red/green/blue (RGB) digital color codes by modulating T from 250 °C to 340 °C for t = 60 s and d = 200 nm (upper panel (a1–a4) in Figure 3c). This was equivalent to an oxide-layer thickness range of 5 to 30 nm (Figure 3d). A different set of colors, from dark green to yellow-brown, was modulated by adjusting t from 50 s to 80 s at T = 350 °C for the same Cu thickness (d = 200 nm) (lower panel (b1–b4) in Figure 3c). This adjustment in the annealing time corresponds to an approximate copper oxide film thickness of 50 to 70 nm. RGB spectra were identified by simulating the reflectance spectra as a function of the $Cu_2O$-layer thickness for the given color samples, and by performing the reverse simulation (Figure 3d and Figure S10, Supporting Information). The abrupt change of thickness observed between a1-4 and b1-4 in Figure 3d occurred because the oxidation behavior changed abruptly at ~350 °C, as shown by thermogravimetric analysis (Figure S11, Supporting Information).

The three samples in Figure 1b (specifically, the SCCF, intermediate film, and PCCF) exhibited the commonly formed brown color after oxidation at 260 °C for 1 min; the homogeneity, saturation, and lightness differed slightly among the film types (Figure 3e and Figure S12, Supporting Information). After further aging at 120 °C for 24 h in air, the color variation was remarkable (Figure 3f). A highly stable nature was retained in the SCCF, whereas the intermediate and PCCF samples were severely deteriorated. The color histogram in terms of RGB digital color codes (Figure 3e and f right panels) represents the tonal range from 0 (black) to 255 (white), as well as the number of pixels. The R, G, and B values in the color histogram are distinctly separate from each other, even in the aged SCCF sample. The observed broadening of distribution for the red values is attributed to a slight roughening of the surface or interface. (Figure S13, Supporting Information). In contrast, the intermediate and conventional PCCF samples after aging degraded to a highly inhomogeneous and darker achromatic color. Note that the initial Cu thickness also alters the color, although much less compared with the $Cu_2O$ thickness. Notably, the color change was almost negligible when the remaining Cu thickness exceeded ~15 nm (Figure S9d, Supporting Information). Figure 3g shows seven representative colors enumerated in terms of their initial Cu thickness (200 to 380 nm); notably, precise control over the color with oxidation time and temperature is clearly demonstrated with a reproducibility of the mean variation of RGB values within ± 3%. ((Figure S14, Supporting Information).

We next investigated whether spatially confined control of oxidation of the SCCF could be realized using laser irradiation. A SCCF sample was mounted on a motorized stage and irradiated using a motorized shutter-equipped continuous laser (wavelength: 488 nm). Notably, absorbance of the SCCF was ~40% (Figure S15, Supporting Information). Reflectance was subsequently measured by a color scientific complimentary metal-oxide-semiconductor camera. Precise control of both laser intensity and duration was required to balance photothermal heating and conductive cooling. The samples were irradiated with a focal spot of 2 μm in diameter ($e^{-2}$) at varying irradiance of 5 to 3500 kJmm$^{-2}$, by modulating the pulse number N (Figure 4a). Immediately after irradiation, colorful multilayered concentric circles were visible in the SCCF, with diameters ranging from 1.7 to 4.5 μm. The circular color pattern exhibited a smooth transition from the center to the surrounding area, suggesting that the depth profile of oxidation was also gradual under irradiation of 2500 to 3500 kJmm$^{-2}$ on the SCCF; notably, only 5 kJmm$^{-2}$ was required for the PCCF. Importantly, the laser dose required to create an oxidation pattern of similar size in the SCCF was ~80-fold greater than that of the PCCF (Figure 4b). The heat-affected zone (HAZ) in the PCCF was ~10-fold larger than that of the SCCF for a similar size of oxidation pattern. These results can be explained by the degree of defects or grain boundaries, which allow for more rapid and longer propagation of oxidation in the PCCF, compared to the SCCF.

After establishing focal oxidation (zero-dimensional, 0-D) as a dot, we investigated whether line (one-dimensional, 1-D) or area (two-dimensional, 2-D) oxidation could be achieved. To deliver the controlled exposure in a predefined pattern, we scanned the sample stage in sync with the laser shutter. Figure 4c and d shows 1-D controlled oxidation in the SCCF by varying the laser power (32–100 kWmm$^{-2}$ with a fixed scanning speed of 200 ms$^{-1}$ and a loop number, N, of 500). The AFM images (Figure S16, Supporting Information) shows the trace of the laser after irradiating 1 and 2.5 min with a fixed laser intensity of 100 kW mm$^{-2}$. The longer irradiation time resulted in appreciable surface roughening due to lattice expansion during oxide formation. Using a raster scan, 2-D patterns were also created, as shown in Figure 4d. The 1-D oxidation lines evident in Figure 4d still have a color gradient, which depends on the minimum pixel size of laser-induced lithography and limits potential applications. The color gradient can be improved by adopting Top-Hat beam shapers.[26] Collectively, our results show that arbitrary patterning of oxidation in the SCCF at near-diffraction-limited precision offers new opportunities for laser-oxide lithography on metal films.

Our results demonstrate that oxidation of the SCCF can be controlled coherently with atomically flat precision. The thickness of the Cu$_2$O layer on the SCCF was manipulated to within an accuracy of 2–3 nm, with a well-defined interface between the oxide and the metal. The colors realized on the SCCF, as demonstrated in our findings, represent the coherence of oxidation front propagation in the Cu lattice, which is a good indicator of oxidation thickness. As an application of Cu oxidation to the photonic-electronic area, we introduce a laser-oxide lithography technique by local oxidation that can be tuned from 1 micron to several tens of microns in diameter for 0-D, 1-D, and 2-D patterns.

## Experimental Section/Methods

Preparation of oxidised Cu thin film: Atomic sputtering epitaxy (ASE) was adopted to grow a pristine single-crystal Cu thin film (SCCF), which is an improved method achieved by modifying the technical limits of a conventional sputtering system. ASE deposition enhances the quality of the metal film and supplies nearly defect-free and grain boundary-free SCCFs. The system was improved by minimization of signal noise originating from grain boundary scattering of electrons in conductors through replacement of the conventional configuration with a single-crystal Cu wiring network.[27-29] Mechanical noises from other equipment, including motors and pumps, also contribute to rough surfaces and defect formation in the films; thus, in our set-up, any vibration caused by ambient noise was minimized to the fullest extent using an anti-vibration system. Anti-vibration techniques during bulk single-crystal growth have been previously introduced.[30,31] In general, lower vibration of the growth system should improve the film crystallinity during thin film growth as well. In reality, minute vibration does not yield noticeable degradation in conventional thin film growth, especially for the polycrystalline thin films. Meanwhile, atomic sputter epitaxy (ASE) aims to realize atomically flat surfaces by stacking atom by atom. Hence, even minute vibration could significantly disturb initial nucleation and lateral growth, especially coherent coplanar merging of nuclei. The ASE helps to provide perfect plasma gases during the sputtering process. Consequently, the Cu film prepared by ASE showed improved RMS roughness, to ~0.2 nm (2Å) under optimal conditions and ~0.3 nm on average [from ~9.1 nm in a previous study].[17]

Oxidation of the Cu film using the aforementioned SCCF was controlled by the following parameters: treatment temperature, treatment time, oxygen partial pressure, and pre-treatment Cu film thickness. The most important factor for homogeneous color is the ability to eliminate the temperature gradient in the sample during the oxidation process. In particular, a heating furnace was designed, which was equipped with a preheating chamber connected to a gas inlet. The temperature gradient in the preheating chamber was <0.1 °C at approximately 300 °C. The sample was inserted into the heated chamber to instantaneously reach the target temperature. The sample was immediately quenched to room temperature after a specific time interval. The treatment temperature of the oxidation process for $Cu_2O$ formation varied from 230–430 °C, according to the target color. The treatment time was occasionally as short as 10 s; in most instances, it did not exceed 5 min. A thick CuO phase (thicker than native CuO layer of > 3 nm, generally forms on the top of $Cu_2O$ layer) emerged when the samples were annelaed > 330 °C for > 5 min. We treated the samples under a mixed-gas atmosphere of Ar (83%):$O_2$ (17%).

Structural characterization: X-ray diffraction (XRD) θ–2θ measurements were performed using a PANalytical Empyrean Series 2 system (Malvern Panalytical, Malvern, UK), equipped with a Cu-Kα source (40 kV, 30 mA). Data were collected within the range of 20° < 2θ < 90°, with a step size of 0.0167° and a dwell time of 0.5 s per point in all instances. Atomic force microscopy (AFM) measurements were carried out using an XE-100 system (Park Systems, Inc., Suwon, Korea). Scanning electron microscopy (SEM), electron backscatter diffraction, pole figure, and inverse pole figure measurements were performed with a Zeiss SUPRA40 VP (Carl Zeiss AG, Oberkochen, Germany), using a scanning electron microprobe. High-resolution transmission electron microscopy (TEM) analyses were performed using an FEI Titan 3 G2 60-300 (FEI/Thermo Fisher Scientific, Waltham, MA, USA), equipped with double aberration correctors (image and probe) and a monochromator operating at an acceleration voltage of 200 kV. TEM samples were prepared with a focused ion beam (Helios 450F1; FEI/Thermo Fisher Scientific).

Optical characterization: Optical reflectance measurements were conducted at room temperature using a near-infrared-visible-ultraviolet (NIR-VIS-UV) spectrometer (Lambda 950; Perkin Elmer Inc., Waltham, MA, USA) over the spectral range of 200–2,000 nm. Reflectance measurements were carried out at near-normal incidence geometry (~10°). A flat aluminum mirror was used as a reference. Optical dielectric functions were characterized using spectroscopic ellipsometry. Ellipsometry spectra were obtained using a rotating-compensator ellipsometer (M-2000; J.A. Woollam, Co., Lincoln, NE, USA) and a rotating-polarizer ellipsometer (VVASE; J.A. Woollam, Co.) at room temperature. A photon energy range of 0.74–5.5 eV was employed, with incident angles of 60°, 70°, and 80°. To obtain the layer structure and dielectric functions of oxidized Cu thin films, we performed optical analyses step-by-step using WVASE software. Figure S7 shows the experimental spectroscopic ellipsometry data and standard fitting procedure. In all procedures, we fitted $\Psi(\omega)$ and $\Delta(\omega)$, which are the ellipsometric angles obtained from the intensity and phase ratio of the reflectance for s- and p- polarized light, respectively, to minimize the mean squared error (MSE). A lower value of MSE means that the fitted spectra are closer to the experimental spectra, indicating higher fitting quality. We began with the known reference complex (real and imaginary) dielectric functions ($\varepsilon_1(\omega)$ and $\varepsilon_2(\omega)$) of $Cu_2O$ and Cu and fitted the thickness of the $Cu_2O$ layer (Figure S7a); the initial dielectric functions of the layers were adapted from Palik.[32]

A thin CuO layer and surface roughness layer were added to improved further the fitting (Figure S7b and S7c) respectively. The layer structure was thus surface roughness/CuO/Cu$_2$O/Cu. Finally, we optimized all parameters together to obtain the result and the actual complex dielectric functions of Cu$_2$O. The final fitted results of the different samples are shown in Figure S8. Consistent fitting of $\Psi(\omega)$ and $\Delta(\omega)$ for the different samples provided reliable $\varepsilon_1(\omega)$ and $\varepsilon_2(\omega)$ values for Cu$_2$O (Figure S8d). Furthermore, the reflectance spectra of oxidized Cu thin films were simulated using the above model layer structure. Change in the reflectance spectra was intuitively understood from the oscillation of color shown in Figure S9d and e, which are reflectance spectra simulations of the sample geometries using the Fresnel equations shown in the inset. The actual fitting of optical functions was necessary only for the Cu$_2$O layer, due to its dominant thickness effect on the optical response of the overall heterostructure. The CuO thickness influenced the color, but the influence was nearly negligible because the native CuO layer was thin without much modulation in thickness. Ellipsometry fitting provided the native CuO thickness of $3 \pm 0.5$ nm for the best fit. This is negligible compared with the controlled Cu$_2$O thickness of 28.3 – 59.9 nm established by optical analyses (Figure S7 and S8).

Setup of laser-oxide lithography and characterization: To build up oxide lithography on the SCCF, the following parts were assembled into a microscope. A continuous laser (LBX-488-500, Oxxius, Lannion, France) with 500 mW power at 488 nm was used to heat the SCCF, given that Cu has a relatively high absorbance below wavelengths of 532 nm. A motorized shutter (SH05, Thorlabs, Newton, NJ, USA) was controlled using software provided by the manufacturer. A 10× objective lens (numerical aperture: 0.45; CFI Plan Apo lambda, Nikon Corp., Tokyo, Japan), a 50:50 beam splitter (BS013, Thorlabs), and a tube lens (TTL200, Thorlabs) for a scientific complimentary metal-oxide-semiconductor (sCMOS) camera (Kiralux CS895CU, Thorlabs) were also included in the lithography system. To oxidize the metal film using the laser, a 2-μm-diameter (1/e$^2$) laser beam was focused onto the SCCF and the power density was varied (200, 400, 800, and 1600 W/mm$^2$) with 100 ms of irradiation. Notably, 100 W/mm$^2$ irradiation was attempted, but clear pattern generation could not be achieved. In addition, laser irradiation with a power density >1600 W/mm$^2$ produced irregular circular shapes, possibly indicating melting of the metal. A supercontinuum broadband laser beam (EXW-12, NKT Photonics, GmbH, Cologne, Germany) was focused on the pattern with a 1-μm$^2$ spot size. The reflected signal was relayed to a spectrophotometer (Semrock SR-303i Spectrograph and Newton DU970-bv EMCCD, Andor/Oxford Instruments). Scanning of the laser pattern was performed using a motorized microscope stage (Ti2-e, Nikon Corp.), controlled by software (NIS-Elements, Nikon Corp.).

## Acknowldgements

This research was supported by Basic Science Research Program through the National Research Foundation of Korea (NRF) funded by the MSIT (Grant Numbers. NRF-2017R1A2B3011822, NRF-2020R1A4A4078780, NRF-2019R1A2B5B02004546, NRF-2019R1C1C1011180, and NRF-2019R1A2C1005267) and Institute for Basic Science (IBS-R011-D1). Spectroscopic ellipsometry has been performed using facilities at IBS Center for Correlated Electron Systems, Seoul National University. The single-crystal target was supplied by the Crystal Bank Research Institute of Pusan National University, Korea. Su Jae Kim, Seonghoon Kim and Jegon Lee contributed equally to this work.

# Figure legends

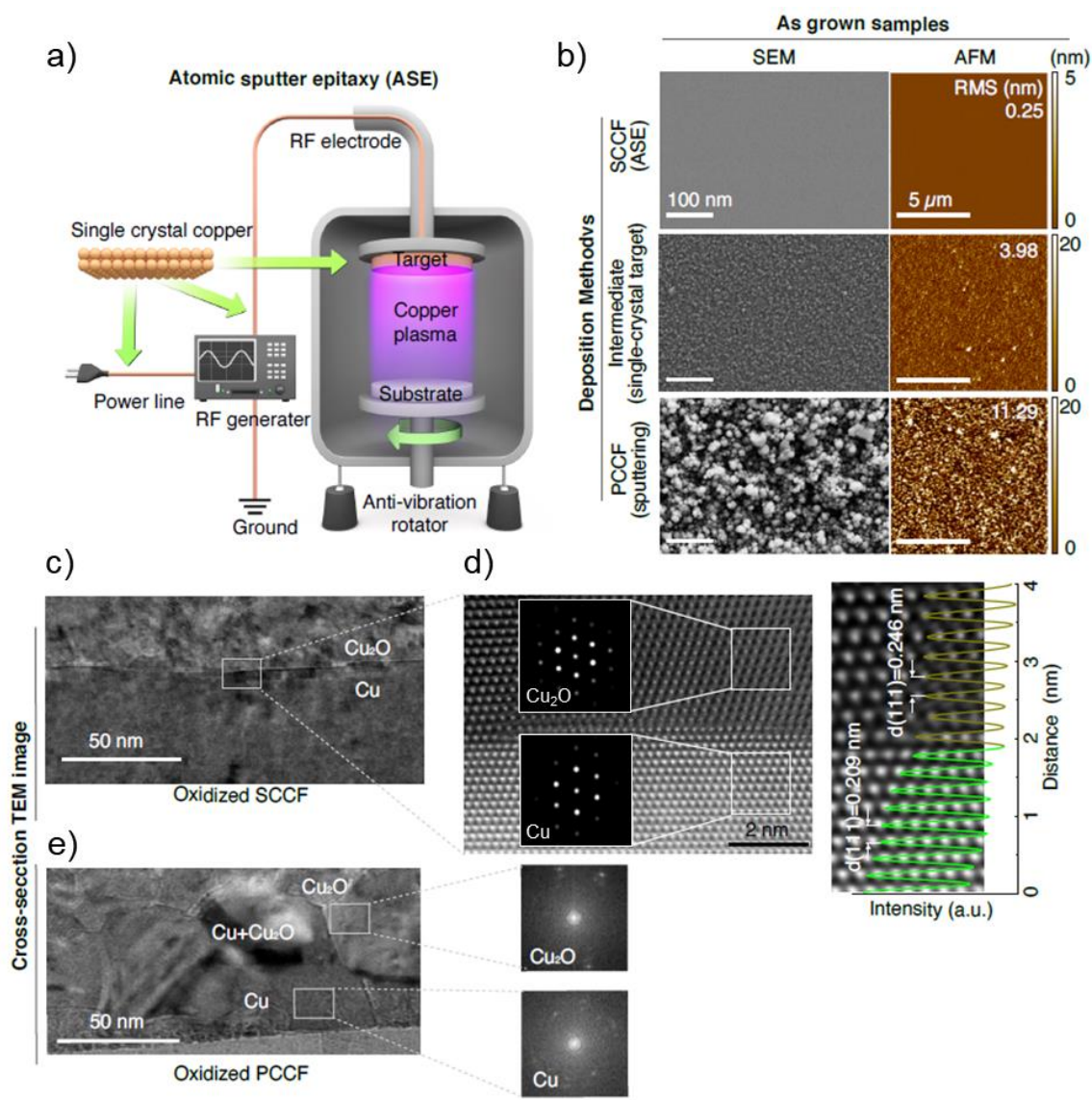

**Figure 1**. Improvement in copper (Cu) film crystallinity using atomic sputtering epitaxy (ASE). a) Schematic diagram of ASE equipped with electrical single-crystal Cu wiring and mechanical noise-elimination system. b) Scanning electron microscopy (SEM, left panel) and atomic force microscopy (AFM, right panel) images of samples obtained from ASE, intermediate-grade film using a single-crystal Cu target, and a conventional sputtering system. c) Cross-sectional

transmission electron microscopy (TEM) image of 50-nm single-crystal Cu thin film (SCCF) (upper) after thermal treatment with an electron diffraction pattern near the interface in the inset. d) High-resolution TEM image of marked area in (c) and a comparison of interplanar spacing profiles between Cu$_2$O and Cu (right panel). e) TEM image of a polycrystalline Cu thin film (PCCF) with corresponding electron diffraction patterns.

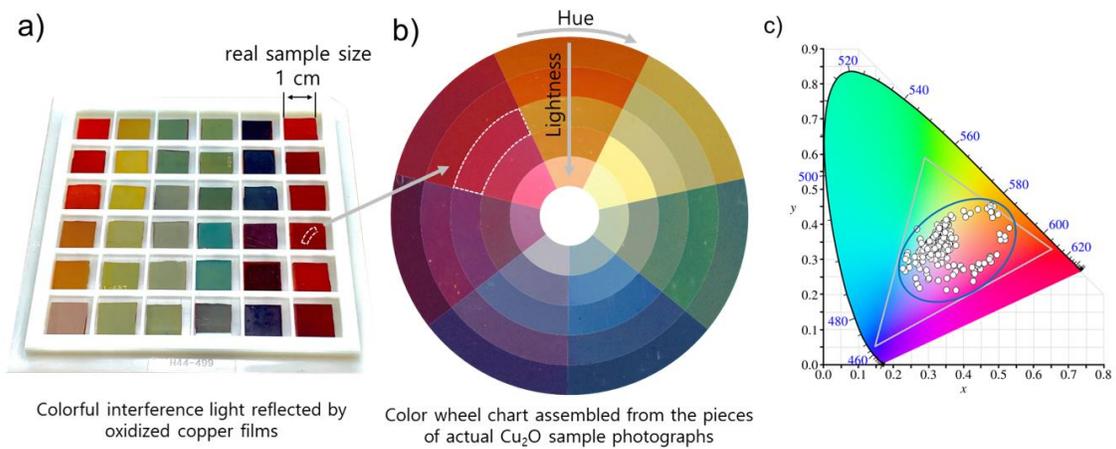

**Figure 2.** Photograph of color maps in single-crystal Cu thin films (SCCFs). a) Photograph of representative SCCFs. b) Color-wheel photograph composed of the representative colors of actual samples. Samples are enumerated with hue in the clockwise direction of the wheel; lightness and saturation are demonstrated in the inward direction. c) Commission Internationale de L'Eclairage (CIE) xy chromaticity diagram with our samples (>300) and reference colors. Gray triangle represents scientific red/green/blue (sRGB) color space of a typical computer monitor. Blue ellipse indicates range covered by the current work, as an area ratio relative to gray triangle.

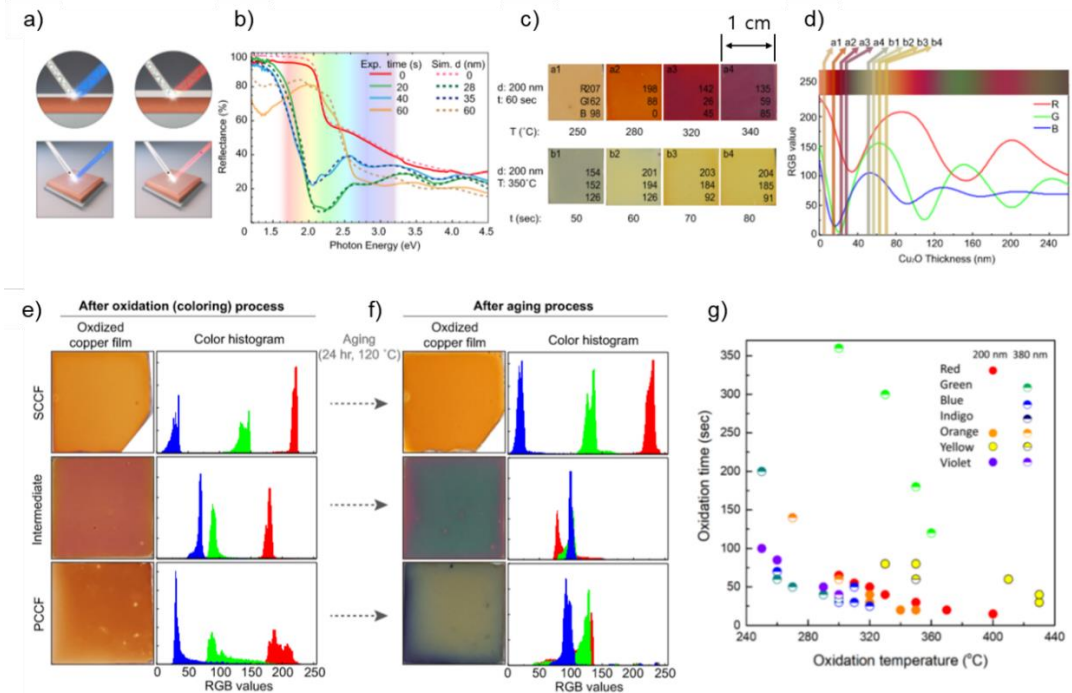

**Figure 3.** Reflectance spectra on Cu thin films. a) Schematic diagram of multiple reflections at optically abrupt interface of $Cu_2O$/SCCF for thin (left) and thick (right) $Cu_2O$ layers. b) Reflectance spectra of SCCF annealed at 330 °C for annealing times of 0, 20, 40, and 60 s (solid lines) and simulated reflectance spectra extracted from the Fresnel equation, rationalizing the mechanism of color control of SCCF (dashed lines). c) (a1–a4) Variation of SCCF (d = 200 nm) by changing T from 250 °C to 340 °C for t = 60 s and (b1–b4) by changing t from 50 s to 80 s at T = 350 °C with red/green/blue (RGB) digital color codes. d) Simulated reflectance spectra of RGB as a function of $Cu_2O$-layer thickness marked by layer thickness for a1–a4 (between 5 and 30 nm) and b1–b4 (between 50 and 70 nm). e) Photographs of colored samples (left) and color histograms (right) after oxidation of samples shown in Figure 1b and f) after further aging for 24 h at 120 °C (of samples shown in panel (e)). g) Seven representative colors with oxidation temperature (T) and duration time (t). Dots and half-solid circles indicate respective thicknesses of 200 and 380 nm.

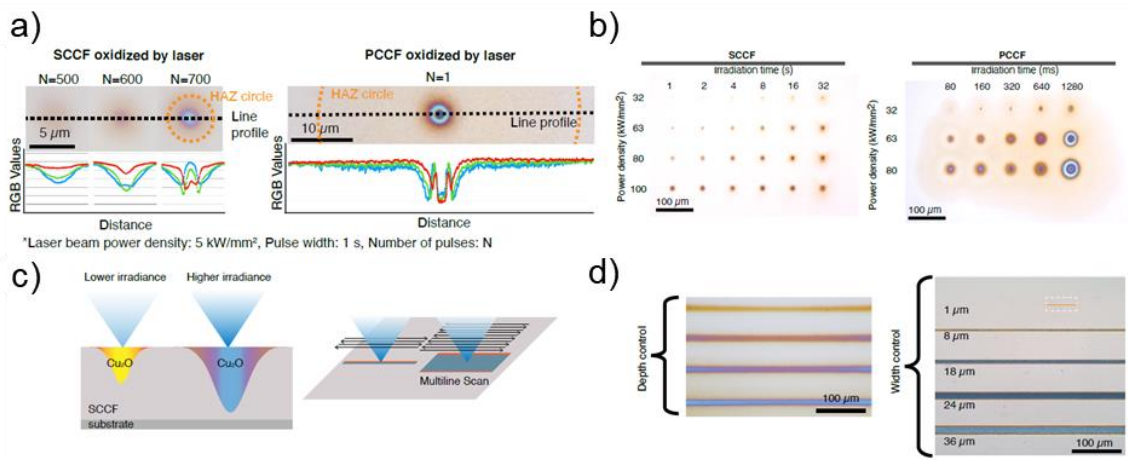

**Figure 4.** Laser-oxide lithography and line profiles. a) Photographs and RGB-color line profiles for focally irradiated SCCFs and PCCFs at varying optical fluence (laser beam power density: 5 kW/mm$^2$, pulse width: 1 s, number of pulses: N). b) Focal oxidation of SCCF and PCCF samples with varying irradiation time and power density. c) and d) schematic diagram and experimental result of optical patterning by laser-oxide lithography with irradiance power. Focal oxidation (zero dimension, as a dot) can be concatenated to form lines (in one dimension) with depth control indicated by coloration, as well as rectangular patterns (two dimensions).

Supporting Information

# Color of Copper/Copper oxide

Su Jae Kim[1,†], Seonghoon Kim[2,†], Jegon Lee[3,†], Youngjae Jo[4], Yu-Seong Seo[3], Myounghoon Lee[3], Yousil Lee[1], Chae Ryong Cho[5], Jong-pil Kim[6], Miyeon Cheon[1], Jungseek Hwang[3], Yong In Kim[7,8], Young-Hoon Kim[7,8], Young-Min Kim[7,8], Aloysius Soon[9], Myunghwan Choi[10], Woo Seok Choi[3*], Se-Young Jeong[7,11*], & Young Hee Lee[3,7,8*]

[1]Crystal Bank Research Institute, Pusan National University, Busan 46241, Republic of Korea.
[2]Research institute of basic science, Seoul National University, Seoul 08826, Republic of Korea.
[3]Department of Physics, Sungkyunkwan University, Suwon 16419, Republic of Korea.
[4]Center for Neurscience Imaging Research, Sungkyunkwan University, Suwon 16419, Republic of Korea.
[5]Department of Nanoenergy Engineering, Pusan National University, Busan 46241, Republic of Korea.
[6] Division of High-Tech Materials Research, Busan Center, Korea Basic Science Institute, Busan 46742, Republic of Korea.
[7]Center for Integrated Nanostructure Physics, Institute for Basic Science, Sungkyunkwan University Suwon 16419, Republic of Korea.
[8]Department of Energy Science, Sungkyunkwan University, Suwon 16419, Republic of Korea.
[9]Center for Artificial Synesthesia Materials Discovery and Department of Materials Science and Engineering, Yonsei University, Seoul 03722, Republic of Korea
[10]School of Biological Sciences, Seoul National University, Seoul 08826, Republic of Korea.
[11]Department of Optics and Mechatronics Engineering, Pusan National University, Busan 46241, Republic of Korea.

* e-mail: choiws@skku.edu, syjeong@pusan.ac.kr, leeyoung@skku.edu

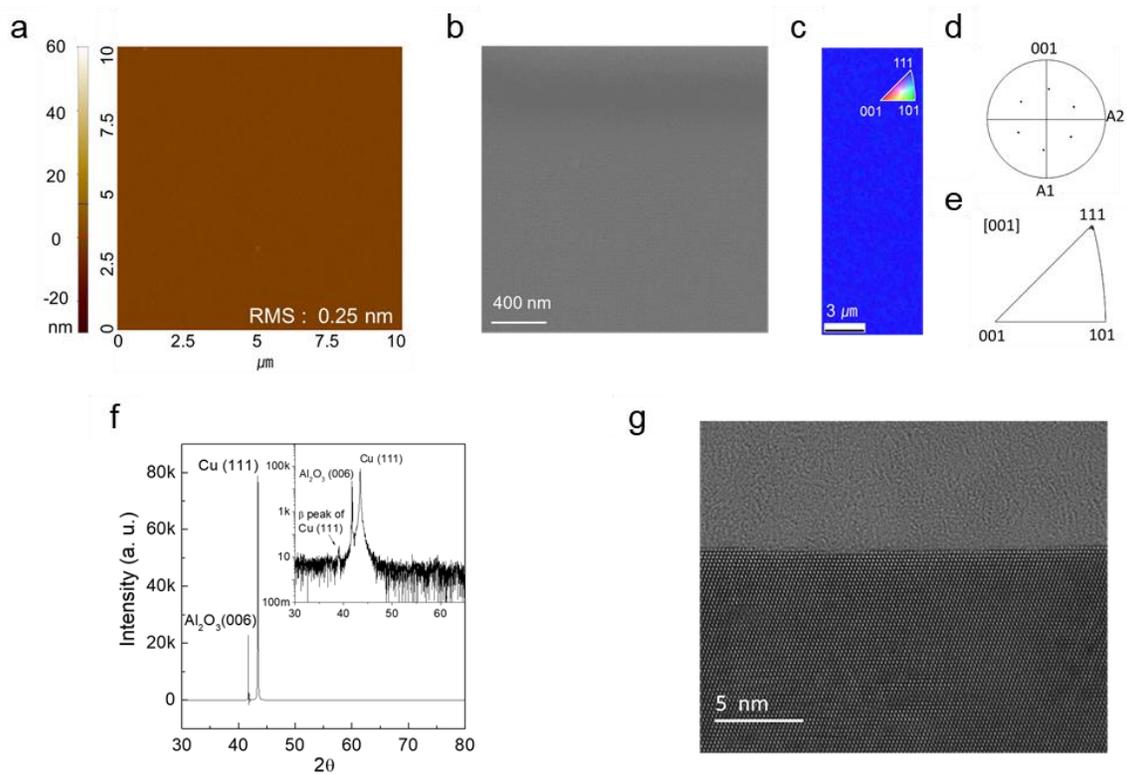

**Figure S1.** Pristine SCCF before heat treatment. a) Surface morphology from atomic force microscopy (AFM) images, with a root-mean-square (RMS) surface roughness of 0.25 nm. b) Scanning electron microscopy (SEM) image exhibiting no grain boundaries at the 400-nm scale. c) Electron backscatter diffraction mapping showing perfect alignment along the (111) plane. d) [100] pole figure showing six-fold symmetry of the {100} plane. e) Inverse pole figure with a sole spot associated with the (111) plane. f) $\theta-2\theta$ X-ray diffraction (XRD) data with a full width at half maximum of 0.060°, comparable to that of $Al_2O_3$ (0.048°). g) Surface of Cu film grown along the [111] direction, with a monolayer-thick step-edge structure observed using cross-sectional (scanning) TEM with 5-nm resolution.

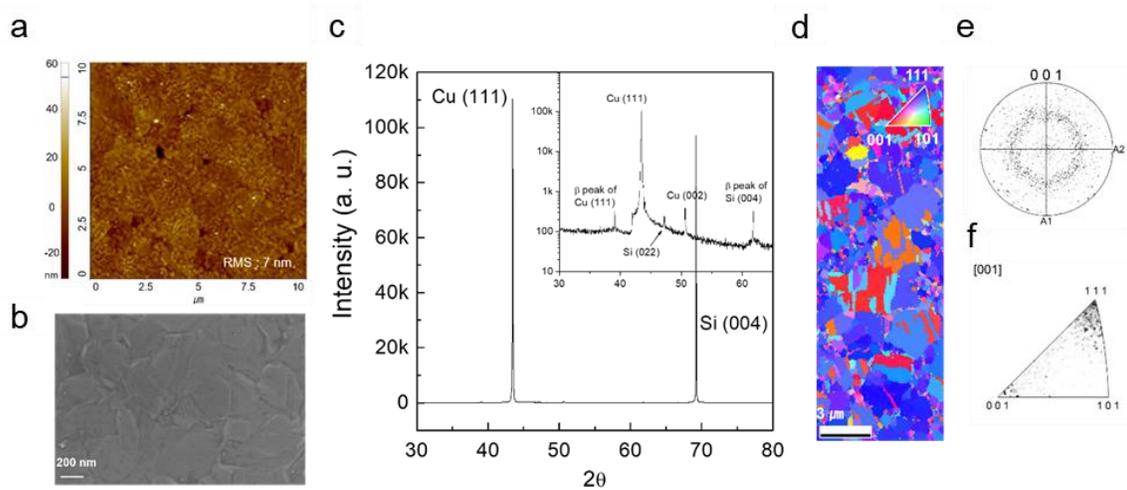

**Figure S2.** PCCF before thermal treatment. a) Surface morphologies from AFM images with a surface RMS roughness of 7 nm. b) SEM image showing grain boundaries of ~$10^9$/cm$^2$. c) θ−2θ XRD data exhibiting mixed phases of (111) and (200). d) Electron backscatter diffraction map depicting numerous grains with different crystallographic orientations. e) [100] pole figure showing various, but less obvious, speckles. f) Inverse pole figure revealing a blurred (111) plane and additional spots.

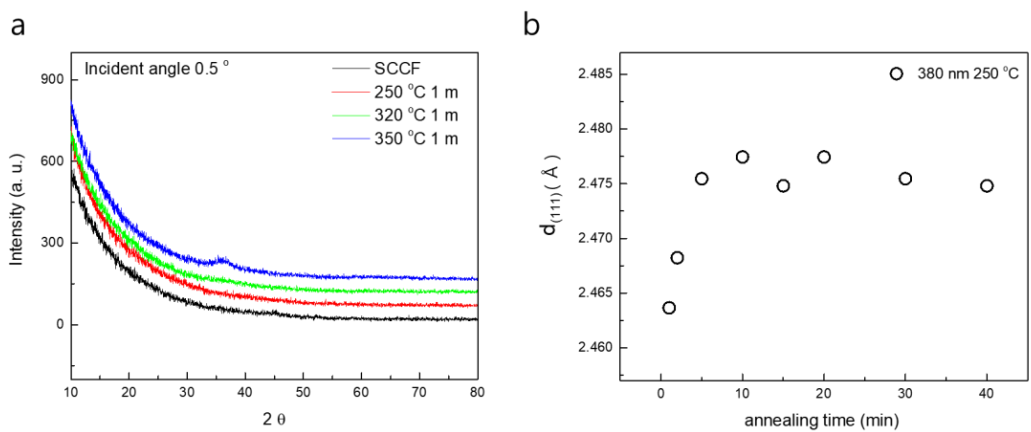

**Figure S3.** a) Grazing incident X-ray diffraction (GIXD) results of films treated for 1 min at different temperatures. b) Change of $d_{Cu2O(111)}$ of the samples treated at 250 °C for different annealing times.

Figure S3a shows the GIXD results at the incident angle of 0.5° for the SCCF annealed for 1 min at 250 °C, 320 °C, and 350 °C. Only the sample annealed at 350 °C showed a small peak near 36°, which reflects that the sample treated at 350 °C has a relatively strained region due to the nonlinear behavior of oxidation above ~ 350 °C. This result shows that that the $Cu_2O$ phase in our color samples treated below 330 °C was uniformly aligned in the (111) direction and had crystallographically well ordered orientation.

Figure S3b shows the change of the Bragg peak position of the sample treated at 250 °C as a function of annealing time to investigate the lattice expansion according to the sample treatment conditions. While 1-minute-annealed sample showed relatively large change of lattice constant due to the influence of the rest Cu, the lattice constant was stabilized with increasing annealing time (increasing $Cu_2O$ thickness). Samples treated longer than 2 min, where the thickness of $Cu_2O$ exceeded ~15 nm, appeared to be relaxed without strain.

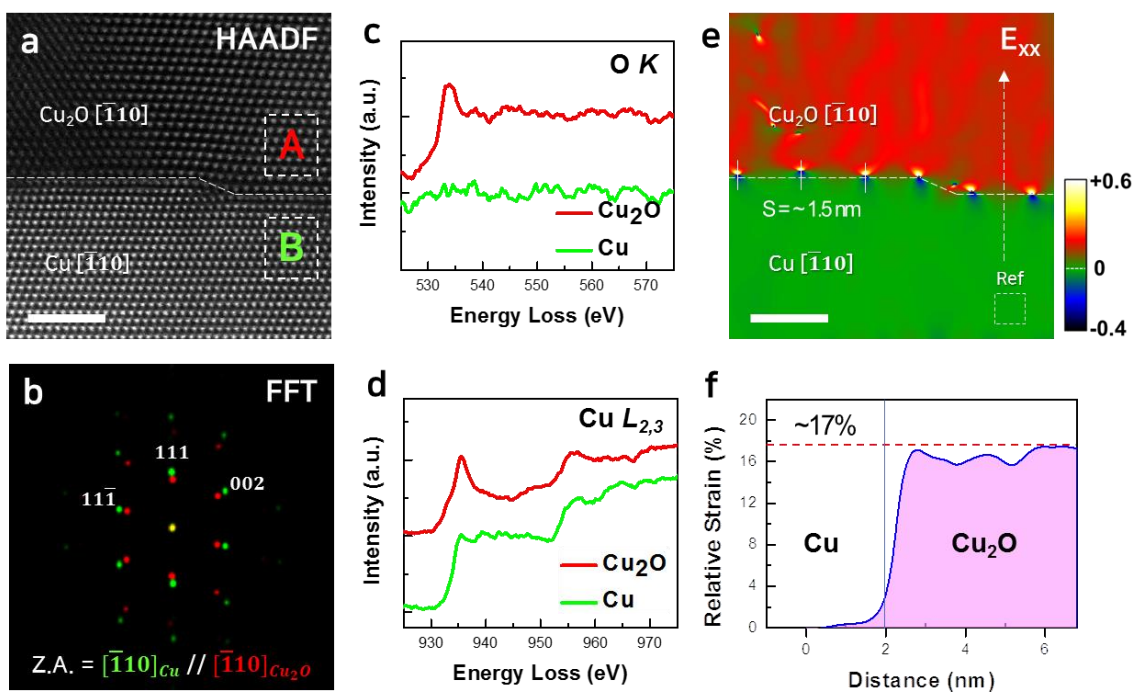

**Figure S4.** Strain distribution across the interface between $Cu_2O$ and Cu. a) Cross-sectional high-angle annular dark-field scanning transmission electron microscope (HAADF STEM) image of the interface of the $Cu_2O$/Cu heterostructure. b) Fast Fourier transform (FFT) pattern of the HAADF STEM image showing the crystallographic orientation relationship of the two crystals as $(111)_{Cu_2O}[\bar{1}10]_{Cu_2O}//(111)_{Cu}[\bar{1}10]_{Cu}$. c,d) Electron energy loss spectra of O $K$ and Cu $L_{2,3}$ edges obtained from the $Cu_2O$ (marked by A, red) and Cu (marked by B, green) layers, respectively. e) In-plane lattice strain map ($E_{xx}$) obtained by geometrical phase analysis (GPA) of the HAADF STEM image, in a). The color scale of the GPA map denotes magnitude of the strain relative to the reference Cu region marked by the white dotted box. f) Strain profile relative to the Cu lattice across the interface obtained along the dotted arrow in e. Scale bars correspond to 2 nm.

Figure S4 shows the strain relaxation behavior of the oxide layer via geometrical phase analysis of the cross-sectional interface structure at the atomic scale. The interface is found to be atomically sharp (*abrupt*) and layer mismatch is noticeably observed at the step edge of the Cu surface. Considering the orientational relationship, the misfit strain ($\delta$) was calculated as ~17%. In this highly lattice mismatched system, it is expected that the in-plane lattice strain is majorly released by repetitive introduction of geometrical misfit dislocation (MD) because lattice mismatch can be more efficiently accommodated by a combination of elastic strain and MDs rather than elastic strain alone. The theoretical average spacing (S) of MDs was estimated as ~1.5 nm according to the relationship of S = (|b|cos30º)/$\delta$, where |b| is the magnitude of the Burgers vector of the perfect dislocation. To understand the microscopic in-plane strain ($E_{xx}$) relaxation behavior of the grown $Cu_2O$ film, we performed geometrical phase analysis (GPA) of the atomic-resolution STEM image.[*Ultramicroscopy* 74 (1998) 131–146] It is evident that the misfit dislocation cores are repetitively placed along the $Cu_2O$/Cu interface and the distance between them is in very good agreement with a predicted spacing of about 1.5 nm. This result clarifies that most of the misfit strain was relieved by the periodic array of the interface misfit dislocations. The interface misfit strain is steeply relaxed over about 2 nm or less by the presence of MDs and the residual strain is gradually relieved over about 4 nm into the $Cu_2O$ layer. (Figure S4f)

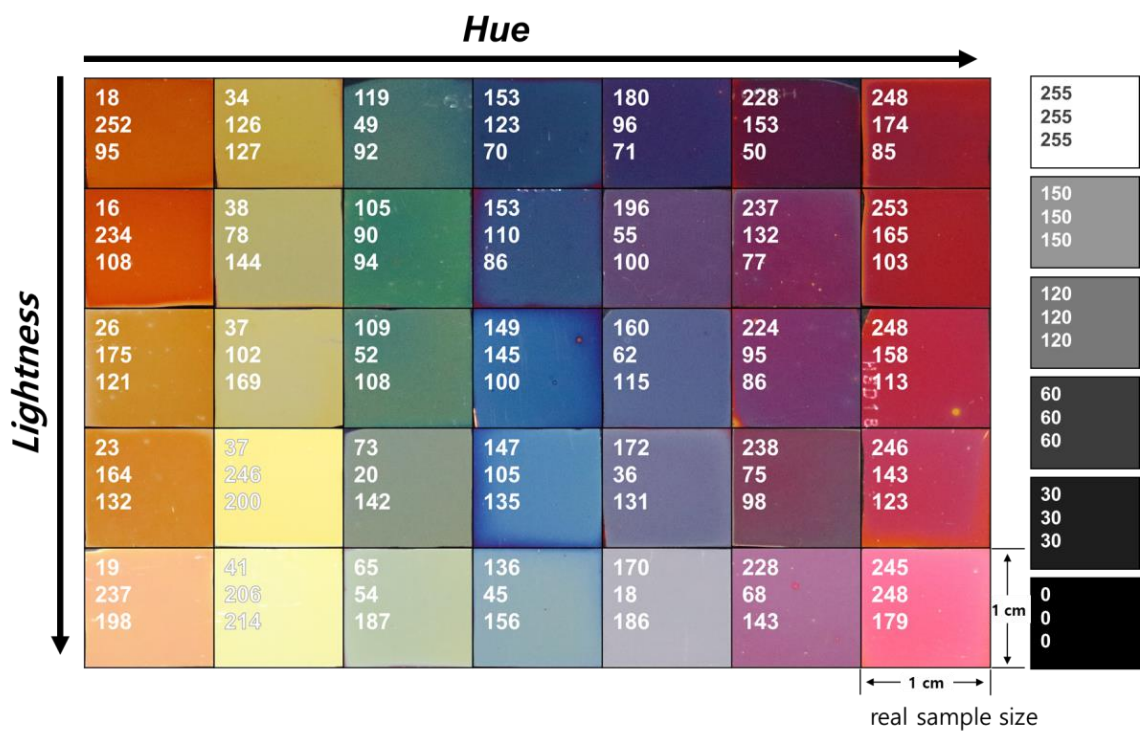

**Figure S5.** Colors implemented on the surface of the Cu thin film. The 35 samples shown in the color wheel of Figure 2b are enumerated along the *x*-axis with increasing hue, and along the *y*-axis with increasing saturation and lightness. Numbers on each sample indicate hue, saturation, and lightness (HSL). For convenience, the HSL ranges (0–255, 0–255, 0–255) follow the standard given in Microsoft PowerPoint. The samples obtained in this study demonstrate different HSL values; however, we selected 35 samples that belonged to a specific color space for better readability. By controlling the parameters of treatment temperature, atmosphere, treatment time, and pristine Cu thickness, we were able to obtain >350 colors. The achromatic colors on the right are provided as visual absolute references.

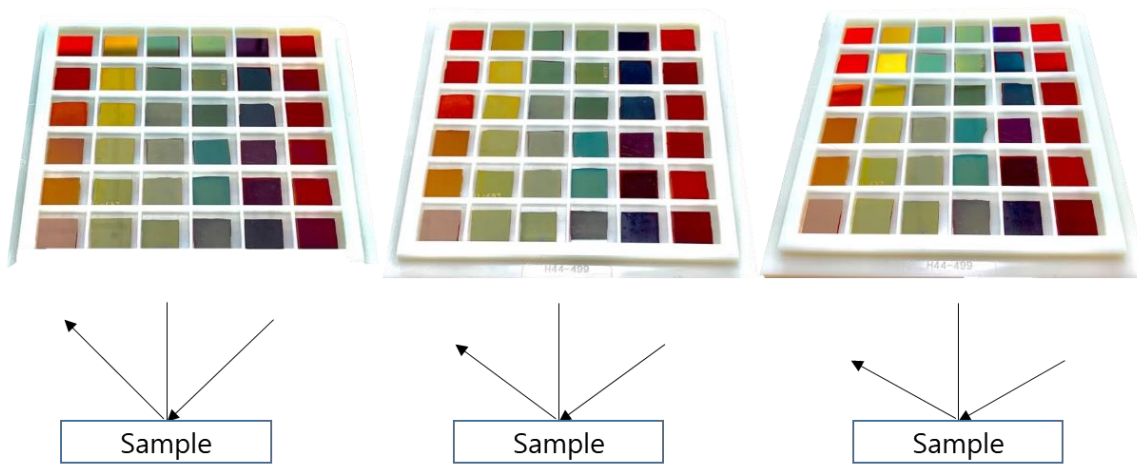

**Figure S6.** Dependence of sample colors on angle of view. The sample colors change slightly according to the different angles of view.

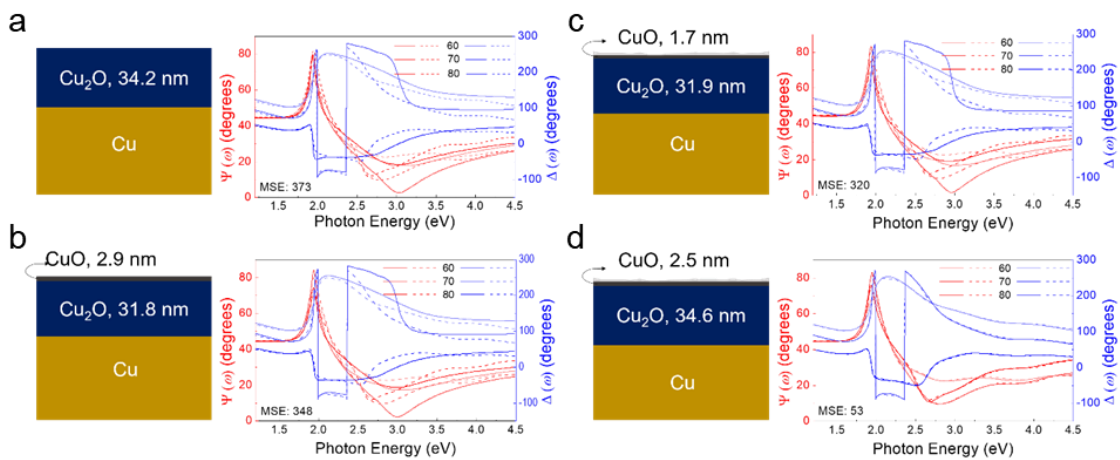

**Figure S7.** Fitting of ellipsometry data. Right panels show comparison between experimental and fitted $\Psi(\omega$, red), and $\Delta(\omega$, blue) spectra for ellipsometry measurements performed at 60°, 70°, and 80°. Straight and dotted lines correspond to fitted and measured data, respectively. Fitting steps are as follows. a) Fitting of Cu$_2$O-layer thickness. b) Simultaneous fitting of thicknesses of CuO and Cu$_2$O layers. c) Adding surface roughness layer (3 nm) with obtained thicknesses of CuO and Cu$_2$O layers. Thus far, optical functions of Cu, Cu$_2$O, and CuO are from the literature.[16,23] d)

Fitting of optical functions of Cu$_2$O layer. During this procedure, the mean-square-error (MSE) was consistently reduced, from 373, 348, and 320 to 53.

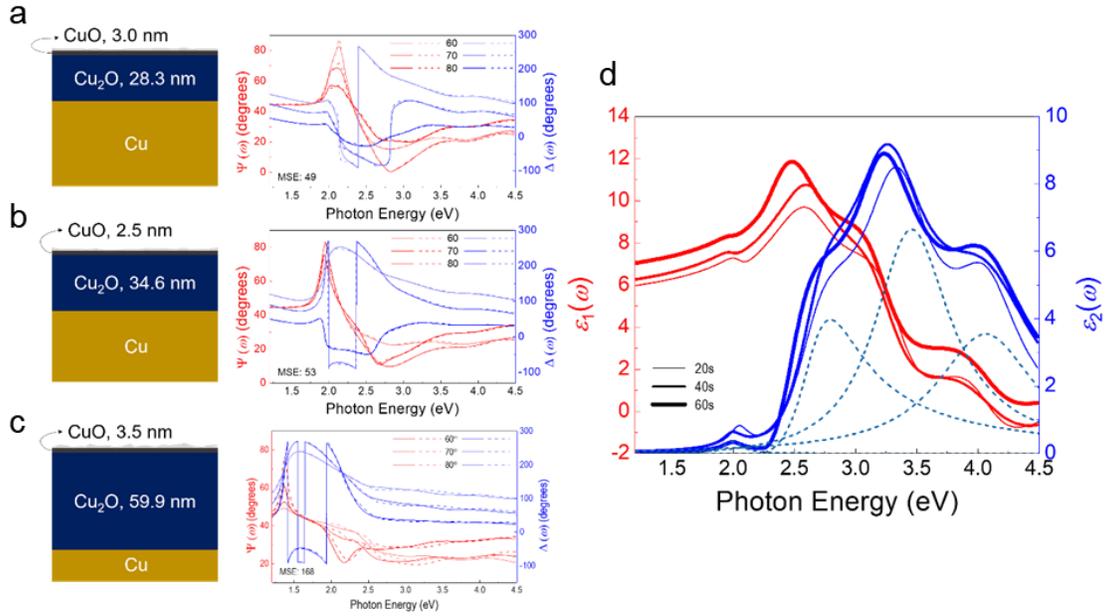

**Figure S8.** Final fitting result of ellipsometry data of CuO/Cu$_2$O/Cu heterostructures shown in main figure. a), b), and c) Right panels show comparison between experimental and fitted $\Psi(\omega)$, red) and $\Delta(\omega)$, blue) spectra for ellipsometry measurements performed at 60°, 70°, and 80°. Straight and dotted lines are fitted and measured data, respectively. Optical properties of SCCFs annealed at 360 °C for 20 s (a), 40 s (b), and 60 s (c) are shown. MSEs of each fitting are 49, 53, and 349, respectively. d) Real [$\varepsilon_1(\omega)$, red] and imaginary [$\varepsilon_2(\omega)$, blue] parts of optical functions of Cu$_2$O layer obtained from spectroscopic ellipsometry measurements. Dotted lines present four Lorentz and Tauc–Lorentz oscillators corresponding to particular optical transitions. Each oscillator represents optical transitions of electrons from valence to conduction state, creating electronic structure of Cu$_2$O.

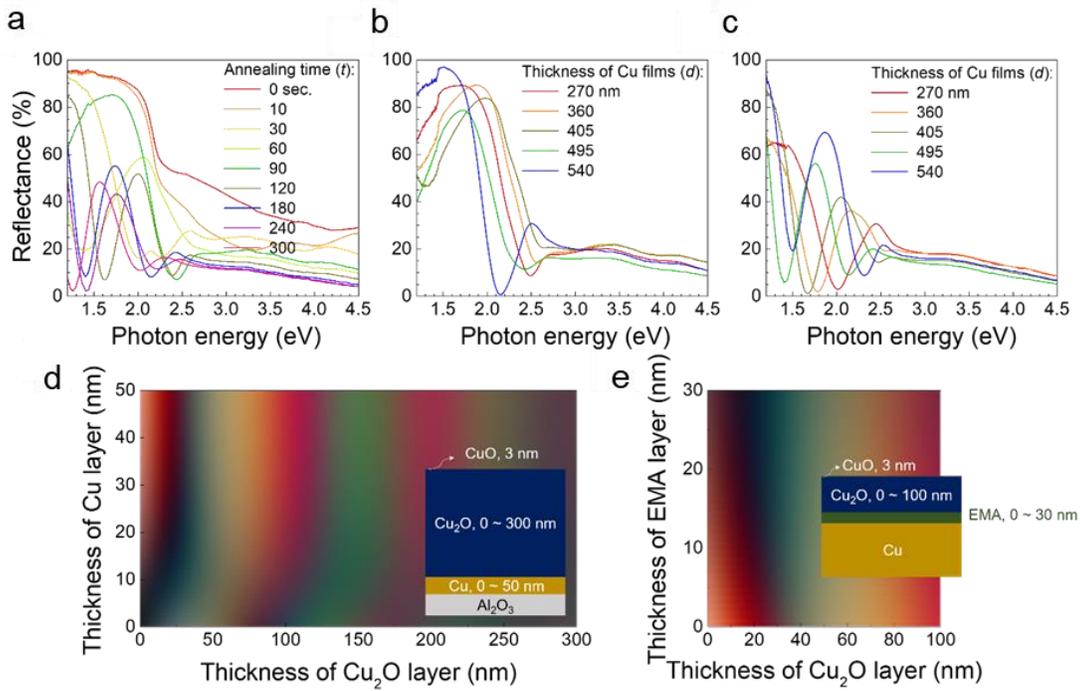

**Figure S9.** Reflectance spectra of oxidized Cu thin films and possible color palette of simulated layer structures. a) $t$-dependent reflectance spectra with $T = 330$ °C and $d = 405$ nm. b) and c) $d$-dependent reflectance spectra with $T = 350$ °C and $t = 1$ min (b) and 2 min (c). d) and e) Simulated color of [CuO/Cu$_2$O/Cu/Al$_2$O$_3$] (d) and [CuO/Cu$_2$O/EMA/Cu] layer structures (e) of various thicknesses, respectively (simulation performed using WVASE software). In the simulation, intermediate layer was constructed using an effective medium approximation (EMA) between dielectric functions of Cu and Cu$_2$O layers; optical functions of each layer were taken from *Palik's Handbook of Optical Constants of Solids*.

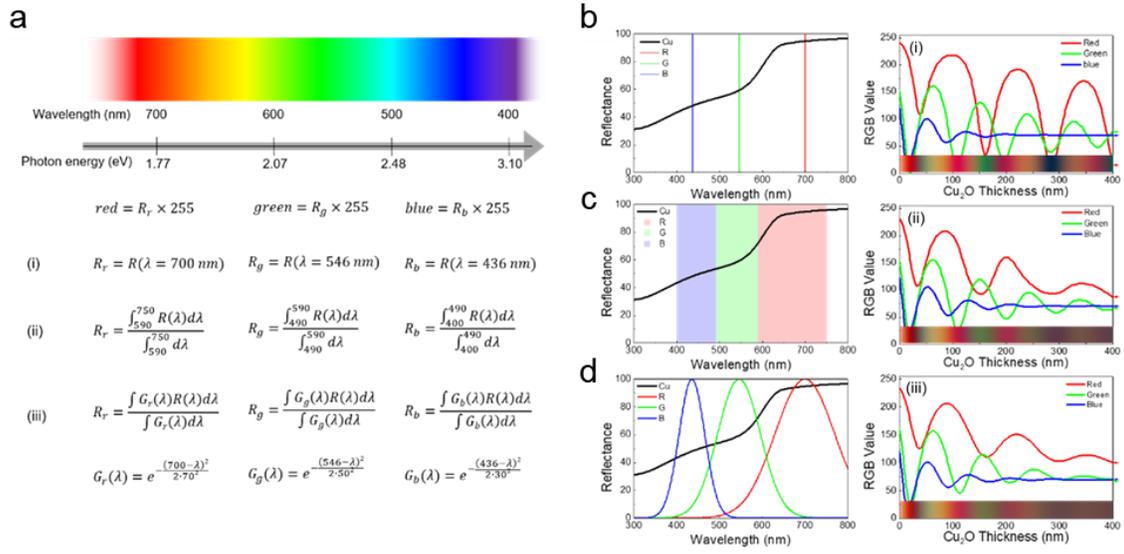

**Figure S10.** Translation of reflectance spectra to RGB color code. a) Analogue color can be simply translated into digital color code using three methods, based on $R(\lambda)$ (i.e., reflectivity as a function of wavelength). (i) Values of RGB color codes calculated by reflectivity at 700, 546, and 436 nm, representing red, green, and blue, respectively. (ii) Values of RGB color codes calculated by normalizing reflectivity across a certain wavelength range. Wavelength regions of 590–750, 490–590, and 400–490 nm were used to represent red, green, and blue, respectively. (iii) Values of RGB color codes calculated by normalizing weighted reflectivity using a Gaussian function. Gaussian functions centered at 700, 546, and 436 nm were used to represent red, green, and blue, respectively. b), c), and d) Each schematic diagram (left panels) and RGB color result (right panels) correspond to methods shown in panels (b), (c), and (d). RGB color result is obtained by simulating reflectance spectra of heterostructure $CuO/Cu_2O/Cu$, with systematic changes in $Cu_2O$-layer thickness. A variety of colors can be realized by changing thickness, indicating effectiveness of oxidation technique. Each color code shows oscillating behavior with different oscillation periods determined by frequency of light.

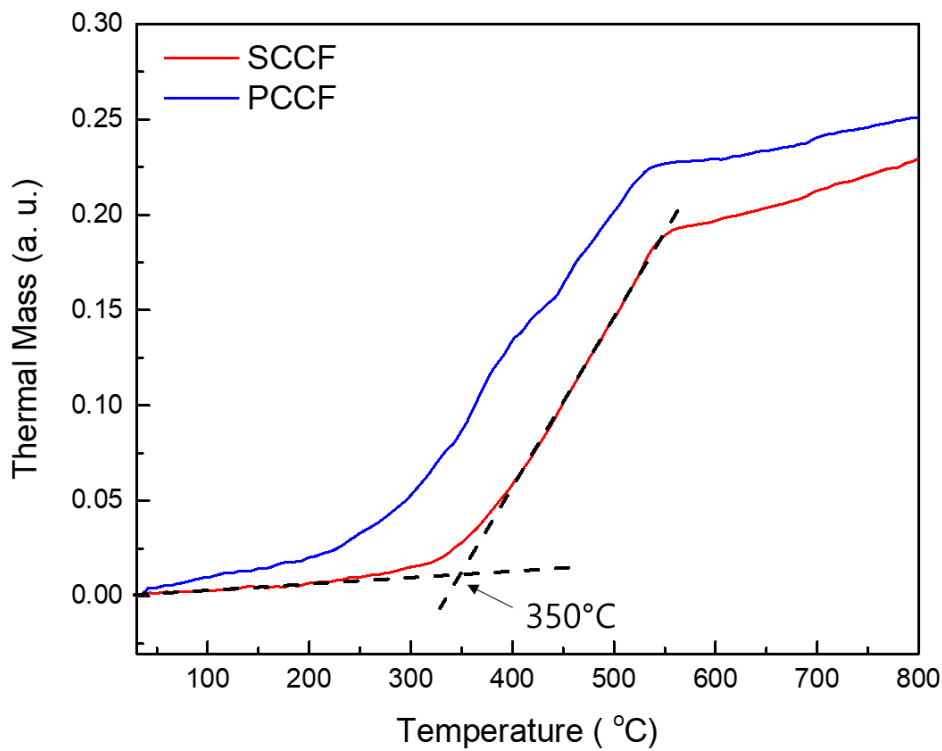

**Figure S11.** Thermogravimetric-Analysis (TGA) result that measured with heating rate 20 °C/min. (SCCF: single-crystal copper thin film, PCCF: polycrystalline copper thin film)

The temperature dependent oxidation behavior of a Cu thin film is rather different from that of the bulk crystal or polycrystalline thin film. Thermogravimetric-analysis (TGA) measured at the heating rate of 20 °C/min (Figure S11) shows that the SCCF thermal mass abruptly at ~350 °C due to oxidation, whereas the PCCF thermal mass increased rather gradually from 220 °C. This result supports the idea that the oxidation behavior of SCCF changes abruptly at 350 °C, in which corresponds to the drastic change of $Cu_2O$ thickness noted by the reviewer.

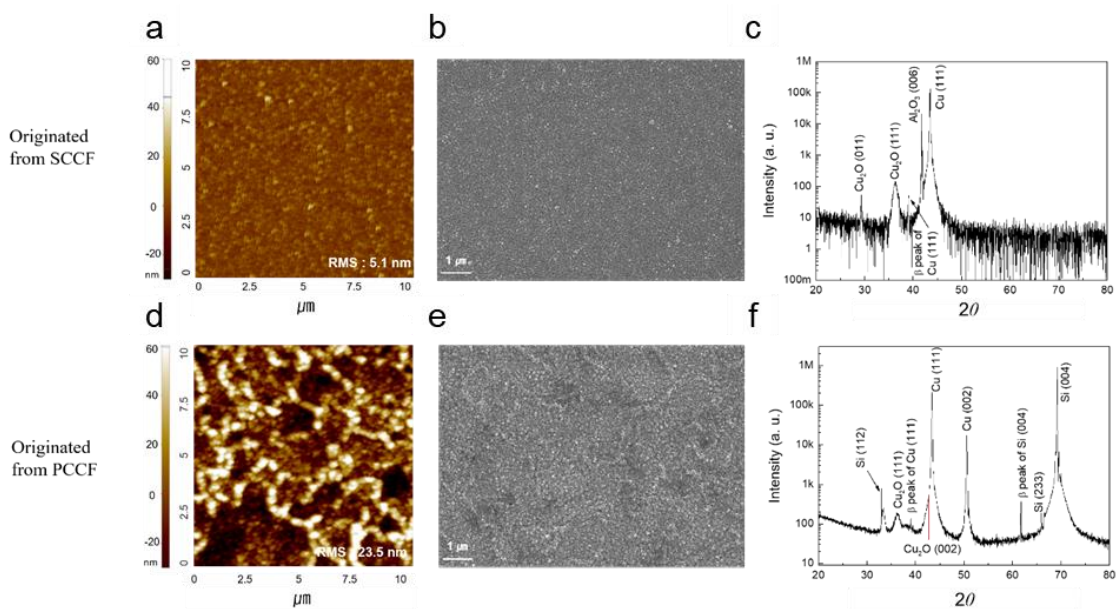

**Figure S12.** SCCF and PCCF after heat treatment. a) Surface morphology obtained from AFM images of SCCF after thermal treatment at 260 °C for 1.5 min; RMS surface roughness was 5.1 nm. b) SEM image of SCCF after thermal treatment. c) θ–2θ XRD data of SCCF after thermal treatment showing high crystallinity along the (111) direction. d) Surface morphology from an AFM image of PCCF after thermal treatment at 260 °C for 1.5 min; RMS surface roughness was 23.5 nm. e) SEM image of PCCF after heat treatment. f) θ–2θ XRD data of PCCF after thermal treatment showing mixed $Cu_2O(111)$ and $Cu_2O(200)$ phases. Figure S12 shows that heat-treated sample from SCCF had epitaxially grown a $Cu_2O$ phase, while the remaining Cu (111) remained unchanged, creating an abrupt interface between oxide and metal where incident light was reflected. However, PCCF with a mixed phase of Cu(111) and Cu(200) was transformed into a mixed phase film of $Cu_2O(111)$, $Cu_2O(200)$, Cu(111), and Cu(200) after heat treatment, where incident light was scattered irregularly.

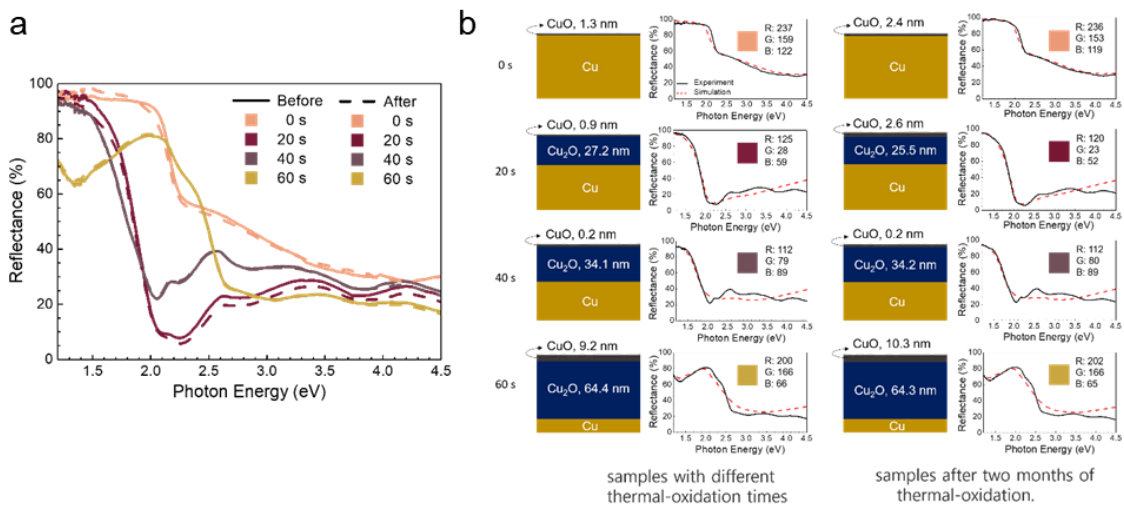

**Figure S13.** Stability of oxidization layer and color of Cu. a) Change in reflectance spectra of Figure 3b over time. Even after 2 months, change in color was minimal. b) Reflectance spectrum fitting indicates that oxidation layer does not become thicker over time under ambient conditions. Fitting was carried out using library $Cu_2O$, with a surface roughness of 1.5 nm.

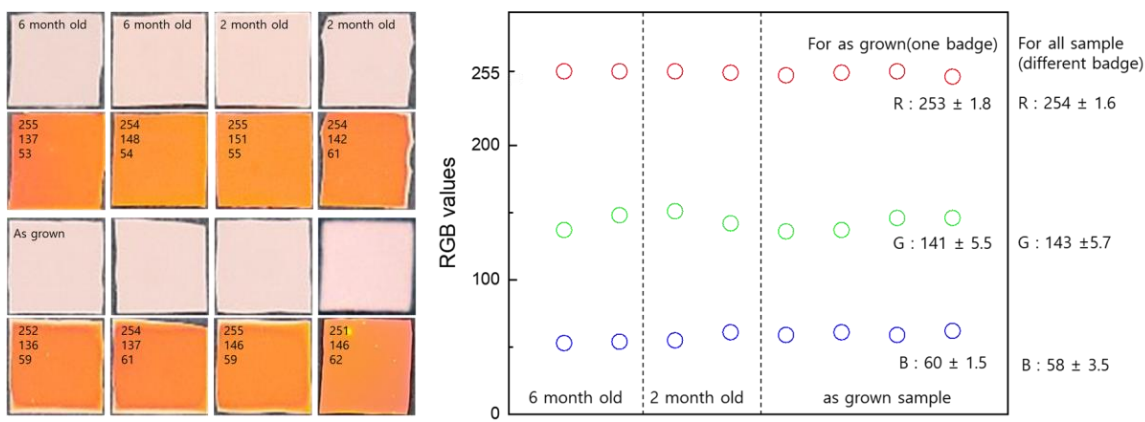

**Figure S14.** Reproducibility of the color from SCCF. The SCCF sample that heat treated at 250 °C, for 2 min for as grown, 2 month old, 6 month old samples.

Coloration of the SCCF was relatively reproducible. Control of the oxide layer thickness was carried out with an accuracy of 2–3 nm, and this tight control resulted in good color reproducibility. Sometimes, depending on the degree of surface roughness of the initial Cu samples, a slight color variation was evident. However, using the same oxidation temperature and time resulted in the same color within the mean variation of the RGB values of ~ 2.8% (± 3.6/255) and the colors obtained from high-quality Cu films displayed consistent color within ~ 2.3% (± 2.9/255). The colors obtained from 2- and 6-month-old and as-grown SCCF samples are presented in Figure S14. While the value of R was relatively invariant, those of G and B changed slightly more than red.

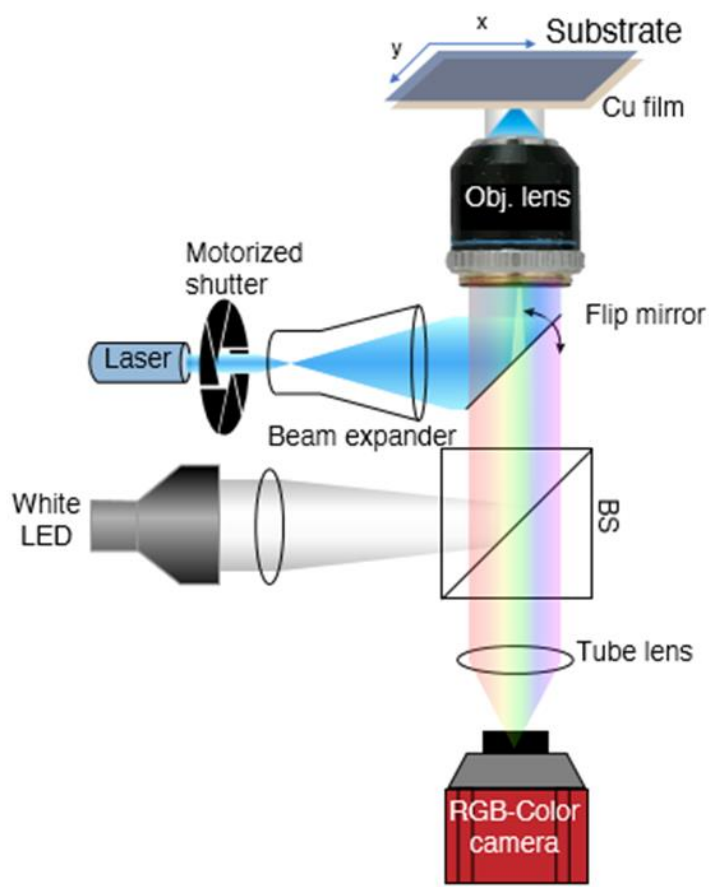

**Figure S15.** Schematic images of sample mounting stage for oxide lithography using a laser. Optical setup for laser irradiation and colorimetric analysis. Cu film is irradiated by a focused 488-nm laser and observed by a color sCMOS camera. BS, beamsplitter; sCMOS, scientific complimentary metal-oxide-semiconductor.

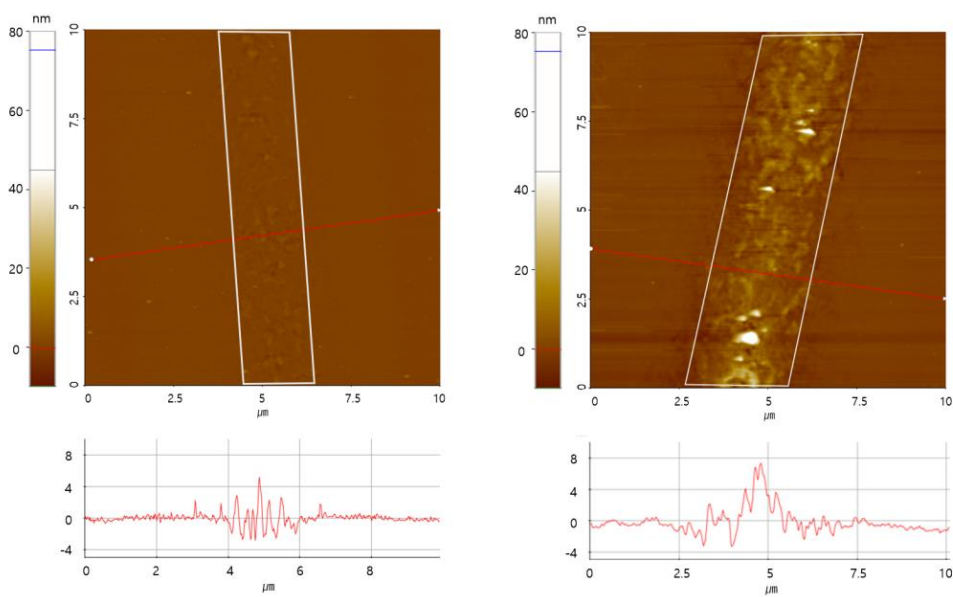

**Figure S16.** AFM image of the colored area after laser irradiation for a) 1 min and b) 2.5 min. The areas marked by white box.

The lattice constant of Cu, $d_{Cu(111)}$ and $Cu_2O$, $d_{Cu2O(111)}$ along the [111] direction is 3.615 Å and 4.26 Å, respectively. The 1.17-times larger lattice constant of $Cu_2O$ means that lattice expansion by oxidation is inevitable.

After irradiating with a laser intensity of 100 kW mm$^{-2}$ for 1 min during a fixed-loop scan (200 m s$^{-1}$ speed, 2 Hz repetition rate), the surface morphology was not appreciably altered. The treated area is marked by a white box in Figure S16a. The line profile below the image indicates a few nanometers of thickening after irradiation. The surface morphology change became more apparent after increasing the irradiation time to 2.5 min. We also observed that the surface height increased with increasing irradiation time (Figure S16b) but the width of the irradiance trace was well maintained. This result supports the idea that coherent oxidation propagation in the depth direction of an SCCF can be achieved by precise control of various laser irradiance parameters, such as intensity, pulse width, repetition rate, treatment duration, wavelength, and beam profile.